\documentclass[%
 reprint,
amsmath,amssymb,
aps,
prab,
floatfix,
]{revtex4-2}

\usepackage{graphicx}
\usepackage{dcolumn}
\usepackage{bm}
\usepackage{subcaption}

\usepackage{graphicx,caption}

\begin{document}

\preprint{APS/123-QED}

\title{Four-Dimensional Emittance Measurements of Ultrafast Electron Diffraction Optics Corrected Up to Sextupole Order}

\author{ M. Gordon\textsuperscript{1}, W. H. Li\textsuperscript{2}, M. B. Andorf\textsuperscript{2}, A. C. Bartnik\textsuperscript{2}, C. J. R. Duncan\textsuperscript{2}, M. Kaemingk\textsuperscript{2}, C. A. Pennington\textsuperscript{2}, I.V. Bazarov \textsuperscript{2}, Y.-K. Kim\textsuperscript{1}, J. M. Maxson\textsuperscript{2} \\
\textsuperscript{1}University of Chicago Department of Physics, Chicago, IL, 60637; \\
\textsuperscript{2}Cornell Laboratory for Accelerator-Based Sciences and Education, Cornell University, Ithaca, NY 14853, USA}

\date{\today}

\begin{abstract}

Ultrafast electron diffraction (UED) is a technique in which short-pulse electron beams can probe the femtosecond-scale evolution of atomic structure in matter driven far from equilibrium. As an accelerator physics challenge, UED imposes stringent constraints on the brightness of the probe electron beam. The low normalized emittance employed in UED, often at the 10 nm scale and below, is particularly sensitive to both applied field aberrations and space charge effects.  The role of aberrations is increasingly important in small probe systems that often feature multiple orders of magnitude variations in beam size during transport. In this work, we report the correction of normal quadrupole, skew quadrupole, and sextupole aberrations via dedicated corrector elements in an ultrafast electron micro-diffraction beamline. To do this, we generate precise  4-dimensional phase space maps of rf-compressed electron beams, and find excellent agreement with aberration-free space charge simulations. Finally, we discuss the role a probe-forming aperture can play in improving the brightness of bunches with appreciable space charge effects.

\end{abstract}

\maketitle


\section{Introduction}

Time resolved tools such as ultrafast electron diffraction(UED), ultrafast electron microscopy (UEM), and X-ray free electron lasers enable the study of non-equilibrium processes in crystals and small molecules occurring at picosecond scales and below \cite{4D-UEDUEM,wang2006potential,SLAC_UED,gao2013mapping,wall2012atomistic,eichberger2010snapshots,emma2010first,XFEL,ishikawa2012compact,LCLS,EXFEL}.  The resolution, both spatial and temporal, of these devices is fundamentally determined by the quality of the electron beam they use.

In UED, the longitudinal degrees of freedom, and in particular the generation of short electron bunches,  have been an important focus of much previous work. Mature longitudinal bunch compression techniques, such as with RF cavities \cite{maxsonprl, luitenprl, daoPRX, zeitler}, THz fields \cite{Kealhofer2016, snively, daoTHz}, or other optical techniques \cite{beatnote, Hassan2018}, have advanced the state of the art to the point that hundreds of femtosecond temporal resolution is common, and single digit fs resolution or below is feasible. 

A growing body of literature in UED/M is focused on improvement to critical transverse  metrics, namely the probe transverse size and divergence, summarized via the normalized emittance \cite{shen2018femtosecond, MEDUSA-Performance-Paper, feist2017ultrafast, houdellier2018development}.  The emittance of the electron beam can increase in transport due to nonlinear fields coming from space charge and electron optics \cite{DowellSourcesOfEmittance, reiser2008theory}.  Across the photoinjector technology presently in use at UED beamlines, space charge forces have typically been alleviated using some combination of precision transverse and longitudinal laser shaping \cite{maxson2015adaptive, maxson2014efficient, penco2014experimental, zhou2007efficient, siqi} to avoid nonlinear fields,  emitting a long aspect ratio bunch with a reduced charge density at the source and employing pulse length compression at high energy \cite{rklinm, cigar, gulliford2015demonstration},  or accelerating the beam with very high gradient to MeV scale energies where space charge forces are relativistically suppressed \cite{wang2003femto, weathersby2015mega, rosenzweig2020ultra, zhu2015femtosecond}.


In contrast to the $\mu$m-emittance dynamics of very high bunch charge photoinjectors for synchrotron radiation applications like x-ray free electron lasers \cite{ecomp}, the nm-emittance dynamics in UED and UEM sources seeking to form small probe sizes \cite{houdellier2018development, feist2017ultrafast, shen2018femtosecond, MEDUSA-Performance-Paper} can exhibit multiple orders of magnitude changes in the transverse beam size during transport in the injector. This is due to at least two factors. The first is that the much smaller emittance is typically achieved by shrinking the source size. Even in the absence of space charge, if the laser wavelength and photocathode material are unchanged, the photoemission divergence will play a  larger role in downstream beamsize for smaller source transverse dimensions. The second factor is that some UED and UEM samples demand small transverse beam sizes owing to sample preparation constraints. Small probe systems, such as ultrafast micro-diffraction techniques, can require comparatively large beam sizes in the final probe-forming lens.  Given that the emittance induced from aberrations scales faster than linearly with the beamsize in the field of the lens, systems with large variations in beamsize will tend to be more susceptible to emittance dilution from aberrations.

\begin{figure*}
    \includegraphics[width=\textwidth]{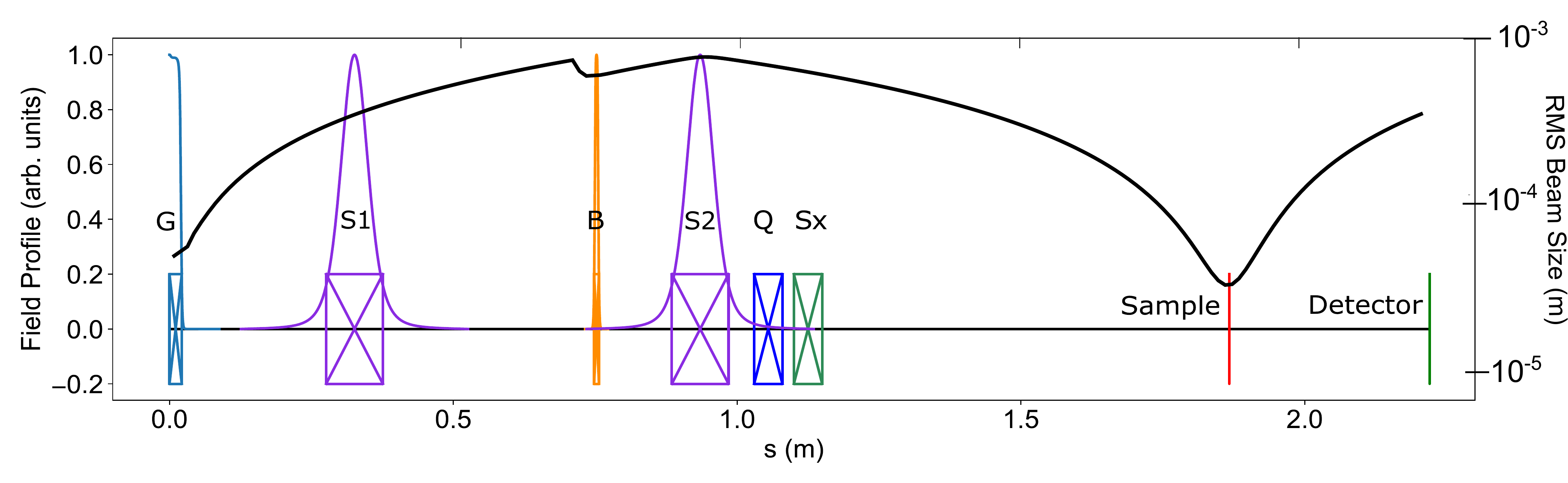}
    \caption{A schematic of the MEDUSA ultrafast electron diffraction beamline, including the 140 keV dc electron gun (G), solenoids (S1 and S2), 3 GHz bunching cavity (B), quadrupole  (Q) and sextupole (Sx) correctors. The plot shows element locations and the normalized longitudinal field profiles (left axis) where applicable. The right axis and black curve shows the evolution of the horizontal rms beam size in logarithmic scale for a typical space charge simulation with 8 fC at the sample plane. The emittances and phase spaces measured in this paper are measured at the sample location. The probe-defining aperture mentioned in text is not included in this simulation.}
    \label{fig:beamline}
\end{figure*}

An extreme example of the these effects is found in conventional transmission electron microscopes which seek sub-angstrom transverse probe sizes. These extreme low emittance beams famously suffer from unavoidable aberrations in electron optics while attempting to maximize the convergence angle in the objective lens \cite{Hawkes2007}. Aberration correctors for these microscopes are extremely complex, featuring multipole elements up to octupole order, but also have been very successful: electron probes are capable of resolving spatial information below half an angstrom \cite{ACTEM-rose1990outline,ACTEM-Rose,ACTEM-Krivanek,ACTEM-Muller,ACTEM-DomainWallFerro}. The beam size evolution in photoinjectors with space charge is far simpler than in electron microscopes. Photoinjectors typically feature only one waist for emittance compensation, in contrast to the multiple crossovers found in electron microscopes. Assuming the aberrations are small perturbations to the divergence (but not necessarily the emittance) at their source, this slow phase advance can allow single corrector magnets placed comfortably downstream to perform near-perfect emittance correction \cite{dowellExactCancellationEmittance2018, Denham2020}.


Quadrupole stray fields in or prior to solenoids couple the transverse phase spaces and dilute the two dimensional emittances. The successful correction of these stray quadrupole fields in solenoid magnets and from rf couplers in photoelectron guns has been well studied and demonstrated with downstream correction quadrupoles \cite{bartnikOperationalExperienceNanocoulomb2015, dowellDevelopmentLinacCoherent2015,dowellExactCancellationEmittance2018, zhengExperimentalDemonstrationCorrection2019, Denham2020}. Previously, sextupole aberrations in solenoid magnets and their correction have been considered theoretically and in simulation, and it was found that a similar correction procedure with a downstream sextupole corrector magnet would be successful \cite{dowellXCorrection}.

In a previous work \cite{MEDUSA-Performance-Paper}, we reported the diffraction performance of the Micro-Electron Diffraction for Ultrafast Structural Analysis (MEDUSA) beamline at Cornell University. We stated that optimal diffraction performance required correction of quadrupole, skew quadrupole, and sextupole aberrations, but did not describe these measurements. In this work we show that our beamline's emittance is sensitive to aberrations up to sextupole order, and describe in detail the correction procedure. To our knowledge, this is the first experimental demonstration of stray sextupole correction in a photoinjector. To diagnose aberrations and beam brightness, we employ a 4-dimensional transverse phase space mapping system with sub-nm emittance resolution. We then describe the brightness performance of the device with rf-compressed bunches containing up to $10^5$ electrons, and show that the measured emittances are in good agreement with aberration-free space charge simulations.

\section{Micro-diffraction optics at the MEDUSA Beamline}
\label{sec:MDO}

The MEDUSA beamline has been described in detail elsewhere \cite{MEDUSA-Performance-Paper}, but we will summarize it briefly here. A schematic of the beamline is shown in Fig. \ref{fig:beamline}. A dc photoelectron gun, here biased at 140 keV, accelerates electrons generated from a Na-K-Sb photocathode grown in our laboratory. The photocathode is driven by 650 nm light generated from a femtosecond optical parametric amplifier; a grating pulse stretcher and pulse stacking crystals create an approximately flat-top distribution with an rms temporal duration of 8.4 ps. Our simulation results presented below best fit our initial photoemission conditions in these measurements to be a transverse initial beam size of approximately $50$ micron rms with a Gaussian profile and $70$ meV mean transverse energy (MTE). Direct measurement of the initial beam size is challenged by the short Rayleigh range (few mm) of our final focus system and the possibility of a small (sub-mm) photocathode recess which cannot be measured directly. This MTE is consistent with near threshold photoemission from Na-K-Sb at this wavelength \cite{maxsontradeoff}. Note that at these sizes and below, the transverse laser shaping techniques mentioned above are extremely challenging given the proximity of the optical diffraction limit.

Red light, rather than the traditionally-used green second harmonic of Yb- or Nd-based lasers, is chosen to reduce the mean transverse energy of photoemission \cite{maxsontradeoff} and thereby to lower the source emittance; more details are discussed in Ref.  \cite{MEDUSA-Performance-Paper}. The pulse repetition rate can be any integer division of 250 kHz. The laser has an angle of incidence of 45 degrees on the cathode which is corrected via the Scheimpflug principle with a rotation of the focusing lens.  An uncorrected spatial asymmetry in the initial distribution can in principle lead to $x-y^{\prime}$ correlations via strong space charge in the solenoids, but in practice, space charge simulations show that in our case, this effect is much smaller than the correlations induced by the stray solenoid quadrupole fields.

After acceleration in the gun, the beam then enters the first transversely focusing solenoid (S1). The beam then is bunched in a 3 GHz continuous wave bunching cavity of the Eindhoven design \cite{luitenprl}, set to produce a longitudinal focus at the sample. After the buncher, a second solenoid (S2) is used to counteract the buncher defocusing and to form the final beam size on the sample. Just downstream of S2, but before the sample, we place a pair of normal and skew quadrupole correctors, as well as one single, rotatable sextupole corrector.  

The main operating mode the MEDUSA beamline is ultrafast electron micro-diffraction with final transverse rms probe sizes on the order of a few microns.  These sizes are generated using a probe-defining aperture \cite{Pietro_Pinhole}, typically 10 micron diameter,  just upstream of the sample. Multiple solenoid setpoints are used in the course of operation to generate various probe charges and coherence lengths as dictated by sample needs. For nearly all operating modes, radiofrequency buncher compression typically generates pulse lengths at the sample $<200$ fs rms \cite{MEDUSA-Performance-Paper}. In this paper we will focus on one subset of setpoints which generate small beam size just prior to the probe-defining aperture.  

A tight focus on the probe-forming aperture is used for samples with low scattering power (such as very thin films) to generate higher charge through the aperture.  We can reduce the spot size on the probe-defining aperture by increasing the beam size in S2, naturally at the expense of a reduction in transverse coherence length. In principle, without space charge, large size in S2 could be achieved by forming a tight waist with S1 prior to S2, but in practice this tight focus is highly nonlaminar and leads to large space charge induced emittance growth. 

The black curve of Fig. \ref{fig:beamline} shows in logarithmic scale a typical setpoint for maximum transmission through the probe-defining aperture. As described earlier, owing to the small initial source size, space charge and intrinsic emittance drive a rapid transverse expansion of the beam in the gun such that at the first solenoid the beam size has increased by more than an order of magnitude. Rather than collimate this beam with S1, in high transmission optics S1 is turned \emph{off} to maximize the beam size in S2. 

In this mode of operation, a practical limitation of the beamsize in the second solenoid arises from the roughly 3 mm diameter through-aperture of the buncher cavity assembly. The clipping effect of this aperture is visible before the buncher field in Fig. \ref{fig:beamline}. Given this aperture,  this mode of operation naturally also has a maximum charge transmittable through the buncher aperture, but in practice this is well beyond $10^5$ electrons/bunch and has not posed a practical limit to operations. Furthermore, for bunch charges at the scale of $10^5$ electrons, collimation or gentle focusing the beam by energizing S1 can in some cases lead to excess nonlinear space charge emittance growth due to increased current density preceding S2.

This optics setting is rather unique in the photoinjector context as by inspection of Fig. \ref{fig:beamline}, the peak beam size is more than 20 times larger than the minimum beam size, and the average beam size from source to detector is more than an order of magnitude larger than its minimum. This makes the beam extremely sensitive to multipole aberrations in both solenoids and the buncher.  Despite this, through the use of quadrupole and sextupole correctors, we are able to simultaneously achieve high transmission and low emittance in agreement with simulation at the sample location.

\section{4D Phase Space Reconstruction}
Our primary transverse beam quality diagnostic is a 4-dimensional phase space scanning technique. We measure the 4-d phase space distribution with our 10 $\mu$m probe-defining pinhole a few cm upstream of our sample location: we focus in the input beam at the pinhole location, and then scan the beam across the pinhole using dipole corrector magnets in both transverse coordinates. For each dipole corrector setting, we capture a two-dimensional image of the transmitted beam on our final detector, a YAG:Ce scintillator screen lens-coupled to a cooled scientific CMOS camera. The intensity distribution of the transmitted beam as a function of dipole corrector setting can be used to generate the spatial distribution of the beam incident on the aperture, and the centroid and width of the electron beamlet transmitted through the pinhole can be used to extract the spatially correlated and uncorrelated momentum widths, ultimately yielding a 4-d density distribution.  Our phase space measurements were performed taking 13 $\mu \text{m}$ spatial step size. The angle resolution, determined by the camera pixel size, image magnification, and the drift distance between the pinhole and the downstream YAG, is approximately 40 $\mu \text{rad}$.

\begin{figure}[htbp]
\includegraphics[width=\linewidth]{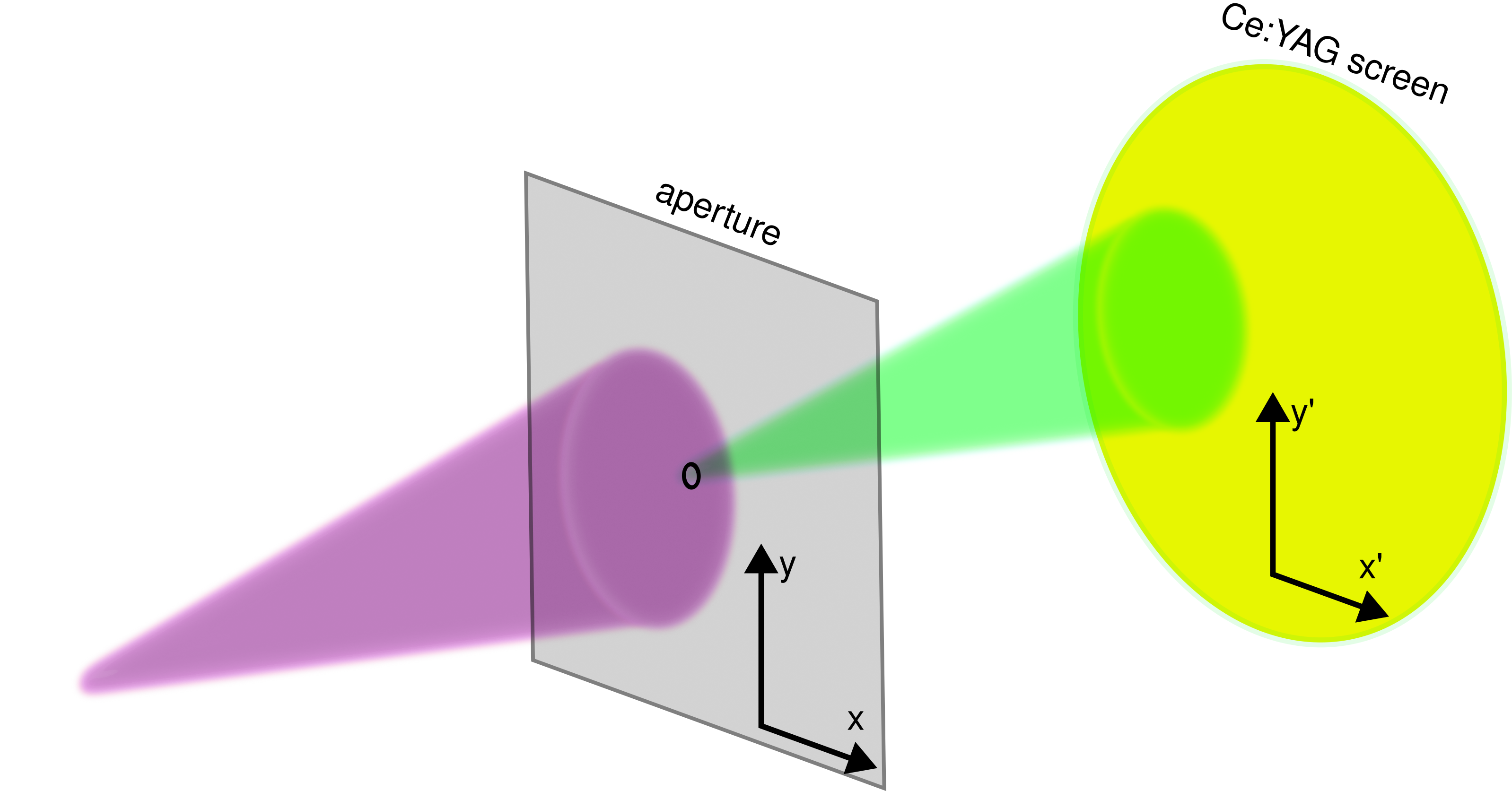}
\caption{A schematic of the method used to measure the four-dimensional transverse phase space density. The incident electron beam (purple) is rastered via a dipole magnet across a 10 $\mu$m diameter aperture in both transverse dimensions ($x$, $y$). An image of each transmitted beamlet (green) is captured on a downstream YAG:Ce screen lens-coupled to a camera, which provides intensity and angular ($x^\prime$, $y^\prime$) information for each beamlet.  }  
\label{fig:aperture}
\end{figure}

With this 4D density matrix information, the 4D beam matrix can be calculated:
\begin{equation}
    \Sigma_{4D}= \begin{bmatrix}
<xx> & <xx'> & <xy> & <xy'> \\
<x'x> & <x'x'> & <x'y> & <x'y'> \\
<yx> & <yx'> & <yy> & <yy'> \\
<y'x> & <y'x'> & <y'y> & <y'y'> 
\end{bmatrix}
\end{equation}
where the prime indicates a derivative with respect to longitudinal position. With the 4D beam matrix, the  4D normalized emittance can be calculated as:
\begin{equation}
    \varepsilon_{4D,n}= (\gamma \beta)^2\sqrt{\mathrm{det}(\Sigma_{4D})}
\end{equation}
To be comparable to 2D measurements of emittance, when values of the beam emittance are presented in this paper, the square root of the 4D emittance will be shown. With these specifications, not only can the transverse plane coupling arising from typical skew quadrupole moments be well resolved, but the transverse emittances can be resolved to well below 1 nm resolution. Our emittance measurement procedure does have the drawback that it is subject to noise from the camera and light pollution from other low intensity light sources. An analysis of our noise removal thresholding procedure shows that the emittances reported here are computed from measured distributions containing at least 90\% of the beam current. To best match this, we also will quote 90\% emittances in all simulations. 


\section{Quadrupole and Skew Quad Correction}
The most important sources of the multipole fields in this setpoint of MEDUSA are the two solenoids and the buncher cavity. In the case of the solenoids, the dominant multipole aberration is quadrupolar.  We suspect this to be caused by the design of the iron yoke of the solenoids, which attach as two separate halves. We believe asymmetry in these two halves to be the primary source of quadrupole moments \cite{bartnikOperationalExperienceNanocoulomb2015}. For the purposes of characterizing the stray quadrupole moments from the solenoids, the first solenoid was energized for the measurements in this section.

\begin{figure}
    \includegraphics[width=\linewidth]{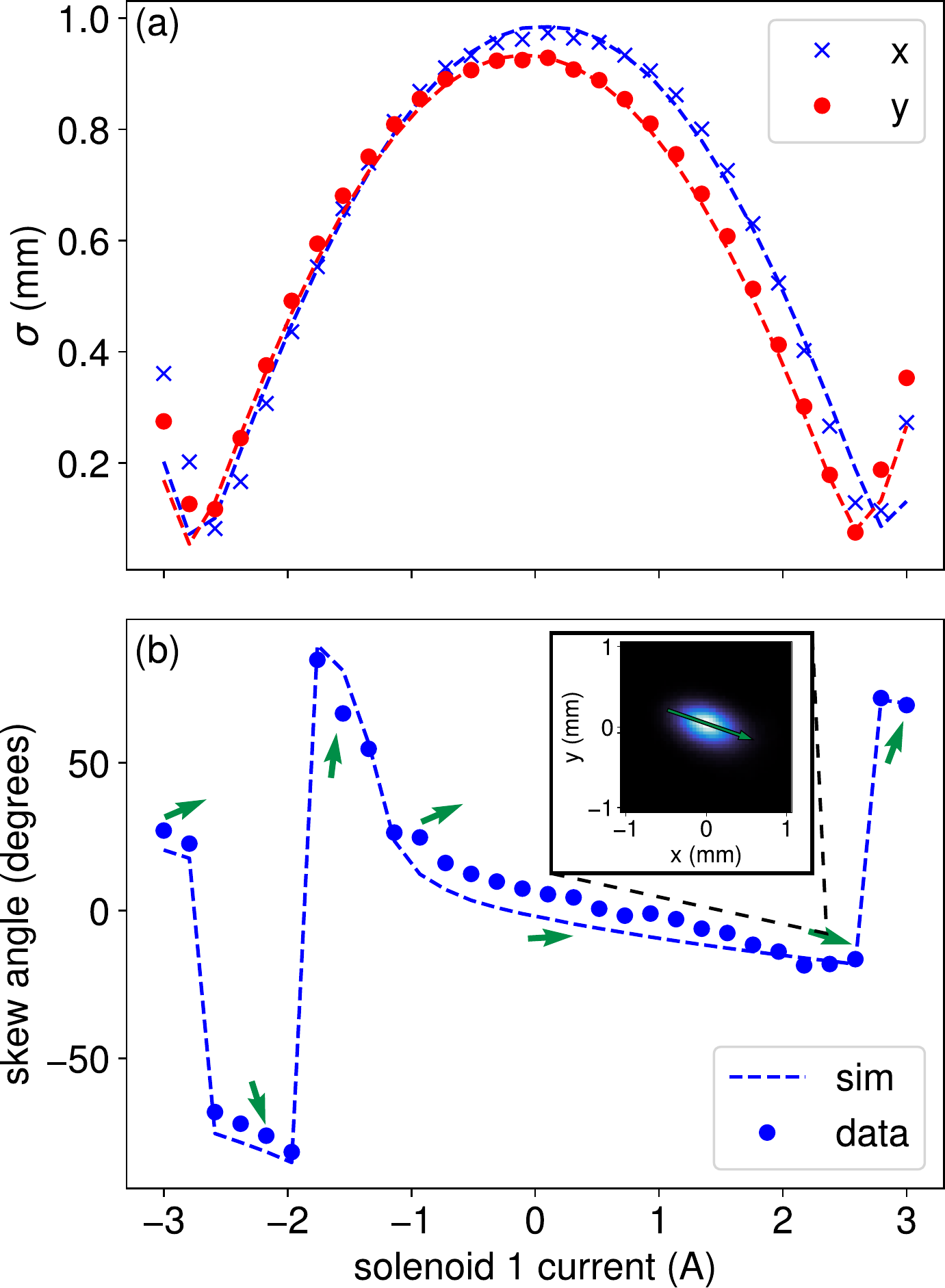}
    \caption{An example of the effects of the stray quadrupole moments in S1 on the (a) transverse beam size and (b) transverse beam shape. The beam shape is quantified with the skew angle of the beam profile. The green arrows provide a visualization of the directions associated with the skew angles. The dashed lines represent fits done with GPT and show good agreement with the experimental data. Inset: Example of a beam with an evident skew. These data are taken on an intermediate viewscreen immediately following S2.  }
    \label{fig:quad_fits}
\end{figure}

Fig. \ref{fig:quad_fits} shows example data of the effects of these solenoid quadrupole moments on the transverse beam profile as a function of the current in the first solenoid with zero field in the buncher. Two effects are immediately apparent: the asymmetry in beam size, particularly near the focus at positive current, and a significant skew angle. We define the skew angle as the arctangent of the slope of the major axis of the beam profile, which causes the discontinuities near the beam focus when the major and minor axes flip. The large discontinuity near -2 A is due to the skew angle being only defined modulo 180 degrees. Using General Particle Tracer (GPT), the particle tracking space charge solver code used throughout this work, we are able to fit for the initial beam parameters and the stray quadrupole moments.

We find that the data is well fitted by modeling two components to the stray quadrupoles moments in each solenoid: a component that scales with solenoid field strength and a constant component arising from hysteresis in the iron yoke. Our model fits the data best when both are oriented as normal quadrupoles located the longitudinal center of the solenoid. 

\begin{figure}
    \includegraphics[width=\linewidth]{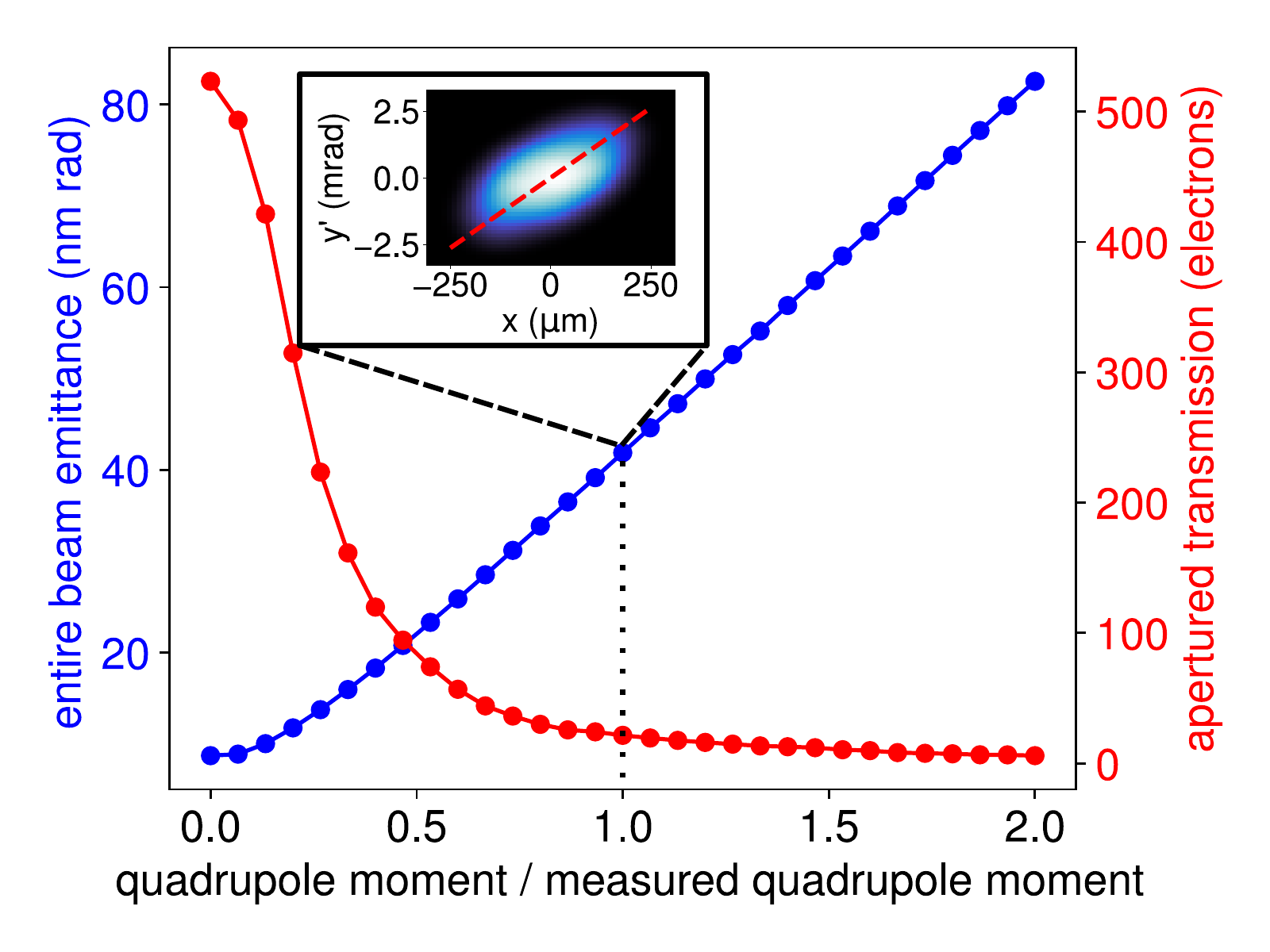}
    \caption{A simulated example of the effects of the stray quadrupole moments on entire beam emittance (blue) and transmission through the final aperture (red). The stray quadrupole moments found by the fits in Fig. \ref{fig:quad_fits} were scaled by a global factor, shown on the $x$ axis, to show the dependence on their strengths. Inset: The $x-y^{\prime}$ phase space measured in experiment. A clear correlation, highlighted by the red dashed line, can be seen.}
    \label{fig:quad_effect}
\end{figure}

The scale of the effects of these quadrupole moments on final beam quality can be seen in Fig. \ref{fig:quad_effect}. Starting with beam settings optimized for maximum transmission through the 10 micron probe-defining aperture (530 electrons, $\sim$ 10 nm rad emittance), the stray quadrupole moments obtained by the fits in Fig. \ref{fig:quad_fits} were introduced in simulation. A tunable global scale factor was applied to the stray quadrupole moments of both solenoids, so that the dependence of emittance and transmission on the strength of the quadrupole could be explored.

The emittance increases roughly linearly with the stray quadrupole strength, while the transmission drops sharply before leveling off. At the fitted strengths, the emittance has increased to 41 nm rad, approximately a factor of 4, while the transmission has dropped to a mere 27 electrons, nearly a factor of 20 lower than optimal. The inset to Fig. \ref{fig:quad_fits} illustrates the large $x-y^{\prime}$ correlation experimentally measured in the presence of these stray quadrupole moments. 

In practical operation, the magnitude of the quadrupole moment introduced in the solenoids depends on both the current setpoint in solenoids and the hysteresis state of the iron, and therefore requires regular tuning when changing optics states. To perform quadrupole correction, a pair of quadrupoles, one normal and one skew, are directly downstream of S2, and are first manually tuned to remove beam tilt and to approximately symmetrize the horizontal and vertical sizes on the final detector, in a procedure very similar to what is decribed in Ref \cite{zhengExperimentalDemonstrationCorrection2019}. Final tuning and verification of the absence of coupling between the $x$ and $y$ phase spaces is performed using a 4-d emittance scan, shown below.


\section{Sextupole Correction}

\begin{figure}[htbp]
\includegraphics[width=\linewidth]{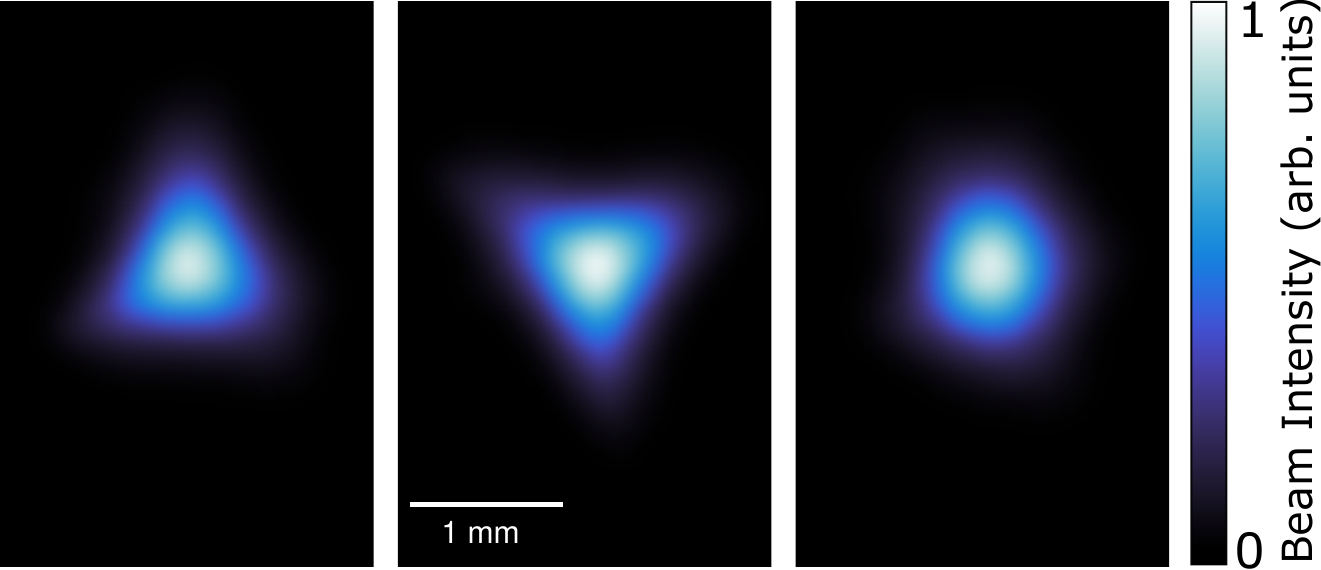}
\caption{Experimental correction of the sextupole moment.  From left to right the beam profiles are taken with: the sextupole corrector off, the sextupole corrector anti-aligned with sextupole moment at a large current, the sextupole correcter anti-aligned at a strength to cancel the sextupole moment. }  
\label{fig:triangle}
\end{figure}

With the buncher and S2 energized, we first perform quadrupole correction as above.  S1 is set to 0 current to minimize the size of the beam in the probe defining aperture, as discussed in section \ref{sec:MDO}. Given the variable state of the quadrupole moment in the solenoids, it is challenging to disentangle what additional amount of skew and normal quadrupole moment are introduced by the buncher. However, after quadrupole correction, the beam shape is triangular on the final detector, shown in the leftmost pane of Fig. \ref{fig:triangle}.  The triangular shape is strong evidence of a stray sextupole field in the beamline. We confirm that the dominant source of sextupole moment is the buncher as the orientation of the triangular distortion can be changed by changing the phase of the buncher, and no obvious distortion exists at zero buncher field.  The buncher is energized via a single input coupler and therefore permits all higher-order multipoles. Note that this suggests also a dipolar and quadrupolar component also arises from the buncher which are corrected as part of orbit and quadrupole correction, respectively, analogous to what has been demonstrated in rf guns \cite{zhengExperimentalDemonstrationCorrection2019}.  To correct for this sextupole moment, a compact sextupole corrector magnet is placed just after S2. 

\begin{figure}[htbp]
\includegraphics[width=\linewidth]{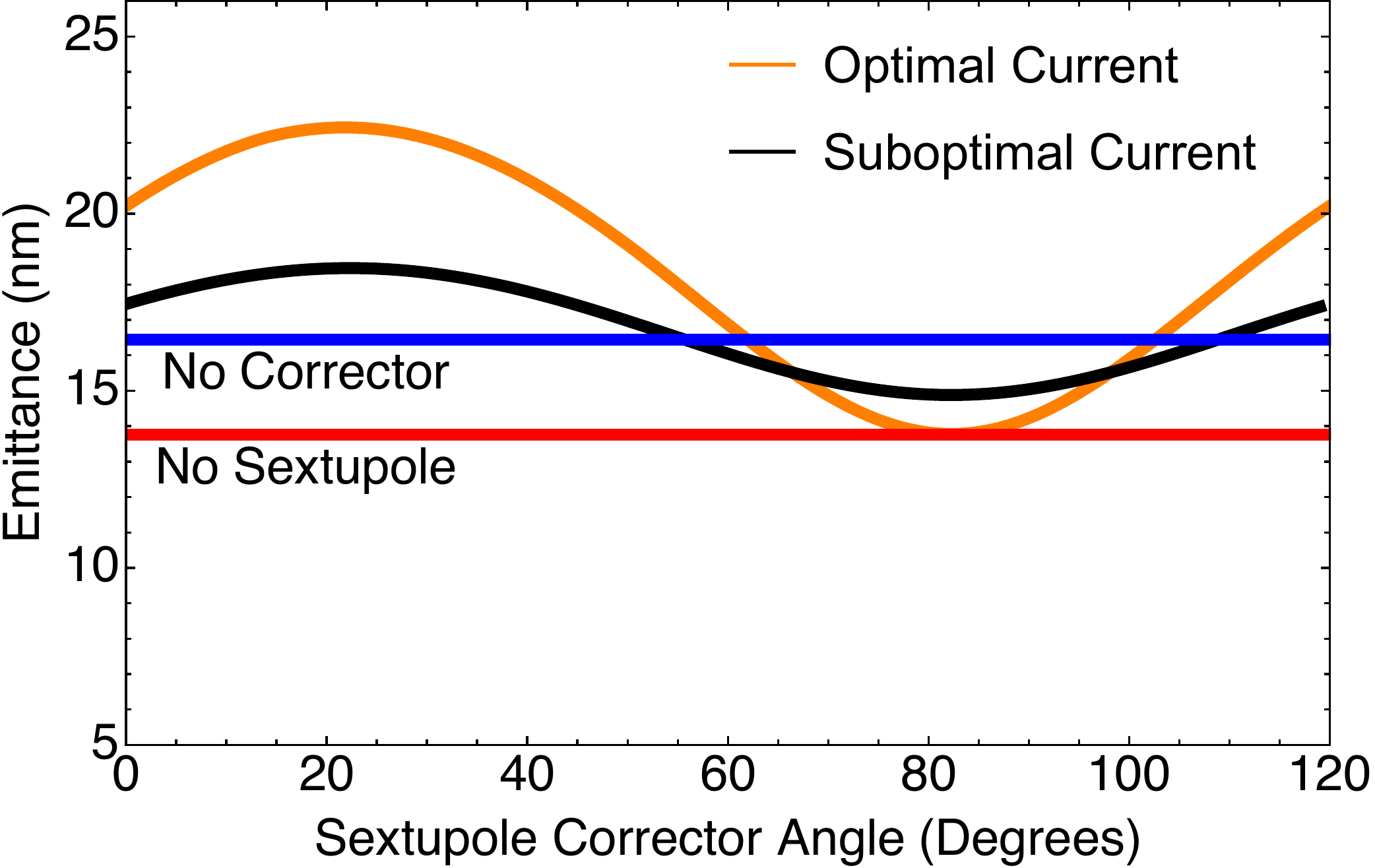}
\caption{Simulated emittance at the sample location varying the angle of a sextupole corrector placed after the second solenoid.  The red horizontal line is the emittance of the beamline with no sextupoles included.  The blue horizontal line is the emittance of the beamline with a sextupole in the buncher and no corrector.  The orange and black curves show the resulting emittance for a corrector current which exactly cancels the sextupole moment and is approximately one third of the needed current respectively.}  
\label{fig:X_Corr}
\end{figure}

Using a downstream sextupole corrector to correct emittance growth from a sextupole in a linear accelerator has been studied previously in simulation \cite{dowellXCorrection}.  Simulations of this effect in our beamline were performed by a placing a thin magnetic sextupole in the buncher and tuning its strength to match the measured emittance and size of beam distortion. This model ignores any potential longitudinal kicks to the beam, and instead is used to approximate only transverse effects. Though the actual distortion of the cavity mode is not a strictly magnetic sextupole, in the thin-lens limit of the cavity, the details of the sextupole kick to the beam are irrelevant, and this is a natural first approximation. The strength of the sextupole scales with the buncher voltage with a fit value of 0.44 T/(m$^2$kV).  A second sextupole was then placed after the second solenoid. In Fig. \ref{fig:X_Corr}, the simulated emittance of the beam at the sample location is plotted against a varying sextupole corrector angle.  It is shown that an appropriate choice of strength and angle in the second sextupole corrects the emittance growth from the buncher sextupole.

To experimentally cancel the sextupole moment, the following procedure is used, which is shown in Fig. \ref{fig:triangle}.  The triangular shape of the beam is recorded with the sextupole corrector turned off.  The sextupole corrector is then turned on to a large current, such that the beam is a triangle with orientation determined solely by the sextupole corrector. This allows us to identify the relative angle between the corrector and the stray sextupole component.  The sextupole corrector is then manually rotated to produce a triangle which is inverted relative to the original distortion.  Lastly the current in the sextupole corrector is gradually reduced until the shape of the beam is no longer triangular. This visual scheme, similar to what is employed for the quadrupole correction, is simple and yields emittance consistent with the existence of zero sextupole moment, as we show below. Interestingly, after sextupole correction, we see the shape of the beam is not perfectly round, and has evidence of even higher order moments.  The magnitude of these moments are relatively small, and we are not sensitive to them in emittance within our experimental uncertainty. Furthermore, while the orientation of the sextupole distortion will depend on the Larmor angle of S2, the variations of beam current are typically small enough in operation that this does not pose a practical problem in our case. For a system with large variations the sextupole angle due to varying solenoid current, an electromagnet with a tunable effective sextupole angle may be required. 

\section{Corrected 4-D phase Space Measurements}

\begin{figure}[htbp]

\begin{subfigure}[htbp]{0.45\textwidth}
\includegraphics[width=\linewidth]{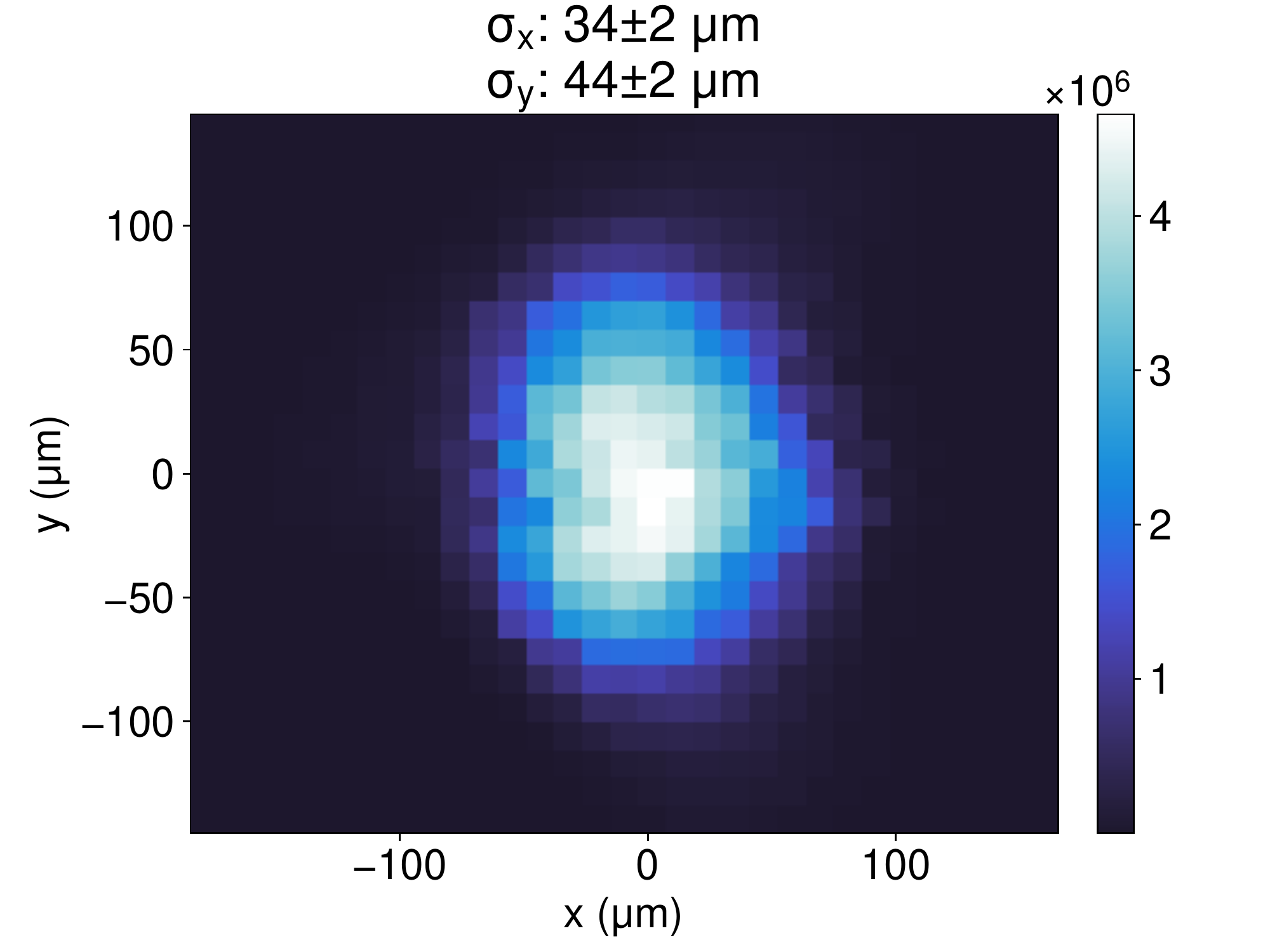}
\caption{}
\label{fig:beam_profile}
\end{subfigure}
\begin{subfigure}[htp]{0.45\textwidth}
\includegraphics[width=\linewidth]{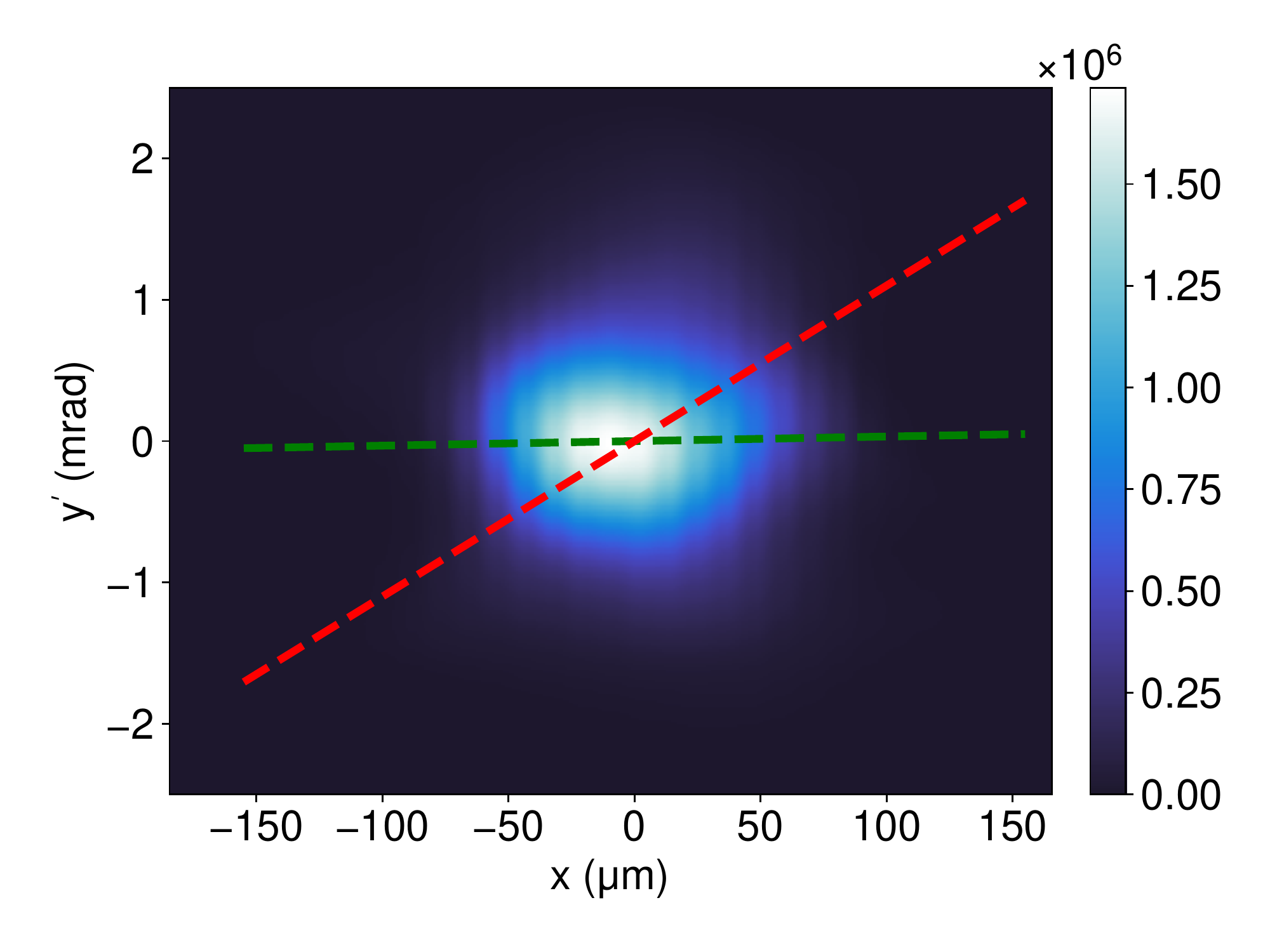}
\caption{}
\label{fig:x_py_phase_space}
\end{subfigure}
\begin{subfigure}[htp]{0.45\textwidth}
\includegraphics[width=\linewidth]{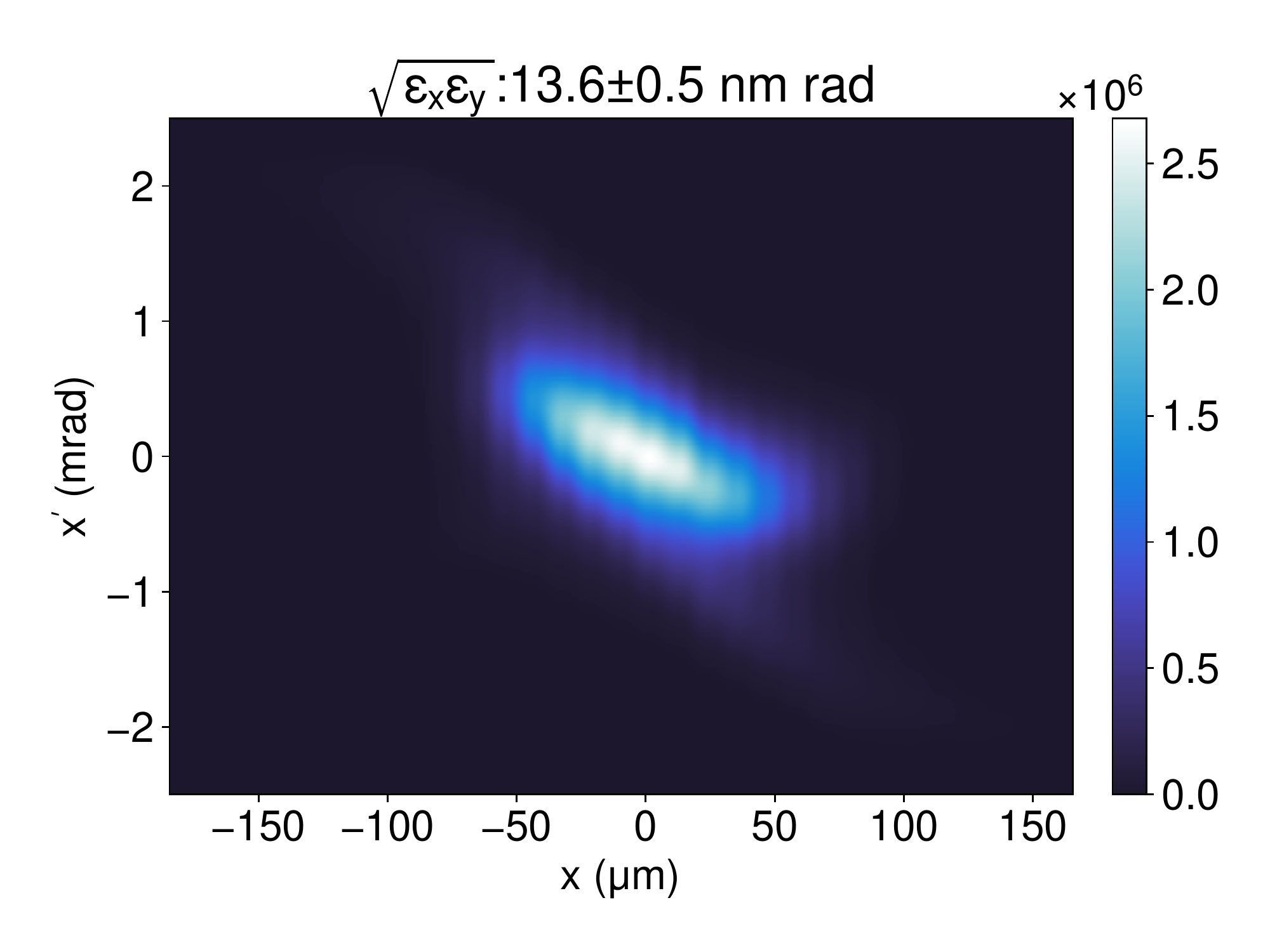}
\caption{}
\label{fig:x_phase_space}
\end{subfigure}

\caption{(a) $x-y$, (b) $x-y^\prime$, and (c) $x-x^\prime$ projections of the reconstructed 4d density matrix at the sample location with sextupole correction.  The dashed green line in (b) shows the correlation between $x$ and $y^\prime$ which is seen to be near 0.  Contrast this to the large correlation in the $x-y^\prime$ phase space in the inset of Fig. \ref{fig:quad_effect}, represented by the red dashed line.}  
\label{fig:4d}
\end{figure}

In Fig. \ref{fig:4d}, we plot the results of 4-d phase space measurements for 8 fC delivered to the probe-forming aperture after quadrupole and sextupole correction. We see in the spatial distribution shown in Fig. \ref{fig:beam_profile} a small asymmetry which may be due to a residual normal quadrupole component. However, Fig. \ref{fig:x_py_phase_space} shows the projection of the phase space distribution onto the $x-y^\prime$ axes. The green line shows the negligible correlation, indicating good cancellation of emittance-diluting skew-quadrupole effects. This is to be contrasted to the red line, reproduced from the uncorrected phase space in Fig. \ref{fig:quad_effect}. Fig. \ref{fig:x_phase_space} shows the $x-x^\prime$ phase space, which shows evidence of space charge effects via a weak, nonlinear tail.

\begin{figure}
\begin{subfigure}[htbp]{0.45\textwidth}
\includegraphics[width=\linewidth]{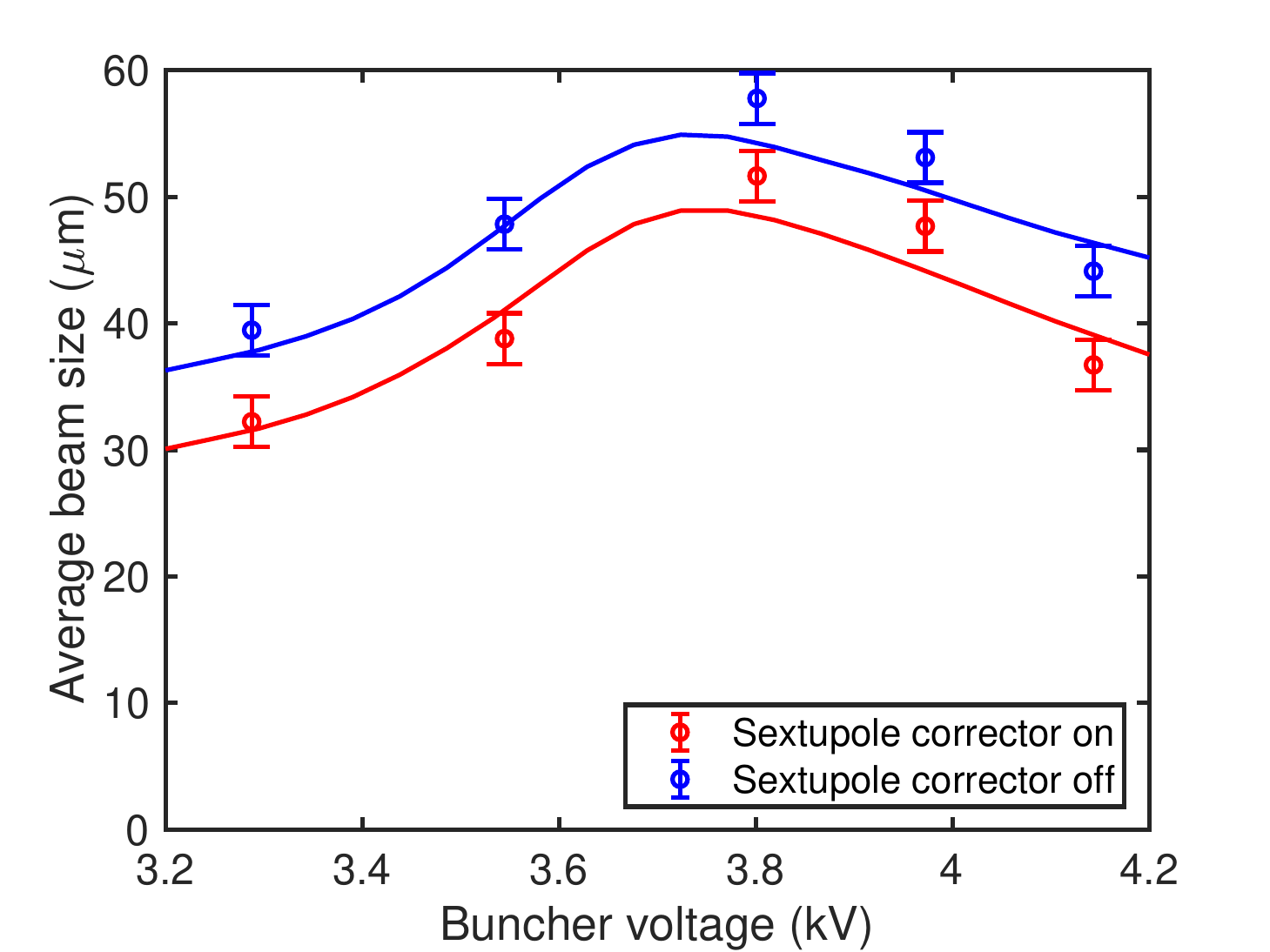}
\caption{}
\label{fig:good_SS}
\end{subfigure}

\begin{subfigure}[htp]{0.45\textwidth}
\includegraphics[width=\linewidth]{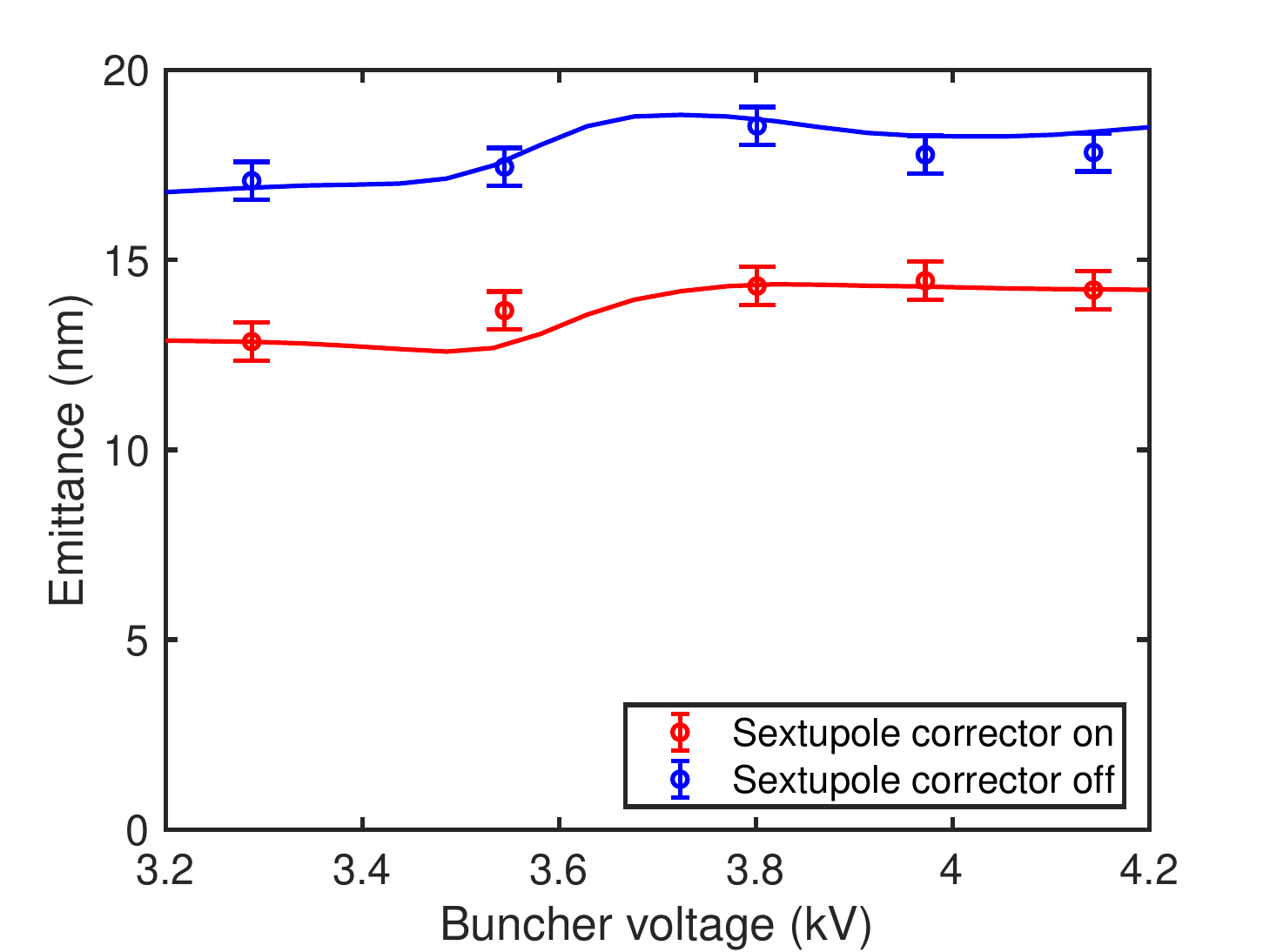}
\caption{}
\label{fig:good_EM}
\end{subfigure}
\caption{(a) Spot size and (b) transverse normalized rms emittance for different choices of buncher voltage around the optimal bunching voltage (3.6 kV). The red line comes from simulation results with zero quadrupole or sextupole aberrations.  The blue line was made simulating the same beamline with a sextupole added inside of the rf buncher, with a best-fit strength of the sextupole chosen.}  
\end{figure}

To verify the cancellation of the sextupole effects, we compare emittance measurements with the sextupole corrector on and off to ``pristine" space charge simulations with zero sextupole moment and to simulations where a thin sextupole of the measured strength is placed in the plane of the buncher.  No quadrupole moments are included inside of the solenoids in these simulations as they are observed to be fully corrected. We perform these measurements as a function of buncher voltage, both above and below the bunching voltage that produces a longitudinal focus at the sample plane (3.6 kV). This also has the effect of modulating the strength of the space charge force near the sample plane, as well as the focusing. For each case, we optimize the strength of S2 to obtain a minimum beam size (and hence maximum transmission) at the aperture. Thus this scan is also a robust test of our space charge simulation fidelity.  These data are shown in Figs. \ref{fig:good_SS} and \ref{fig:good_EM}. Uncertainties in beam size and emittance are chosen based on machine state reproducibility.  The measurements at the operating point for UED, with buncher voltage around 3.6 kV, were repeated 5 times, and the standard deviation of these measurements was used as the uncertainty.  Two free parameters were optimized in the simulation to best match the measured data scale and shape, the cathode MTE and the initial laser spot size.



We find excellent overall agreement with simulation when comparing both transverse spot size and normalized emittance, suggesting the sextupole moment has been well corrected. We also find that for our optics, sextupole correction has a stronger effect on the emittance than the spot size, suggesting that in UED it primarily improves the coherence length of the probe after passage through the aperture, rather than dramatically increasing transmitted charge.

\begin{figure}[htbp]
\includegraphics[width=\linewidth]{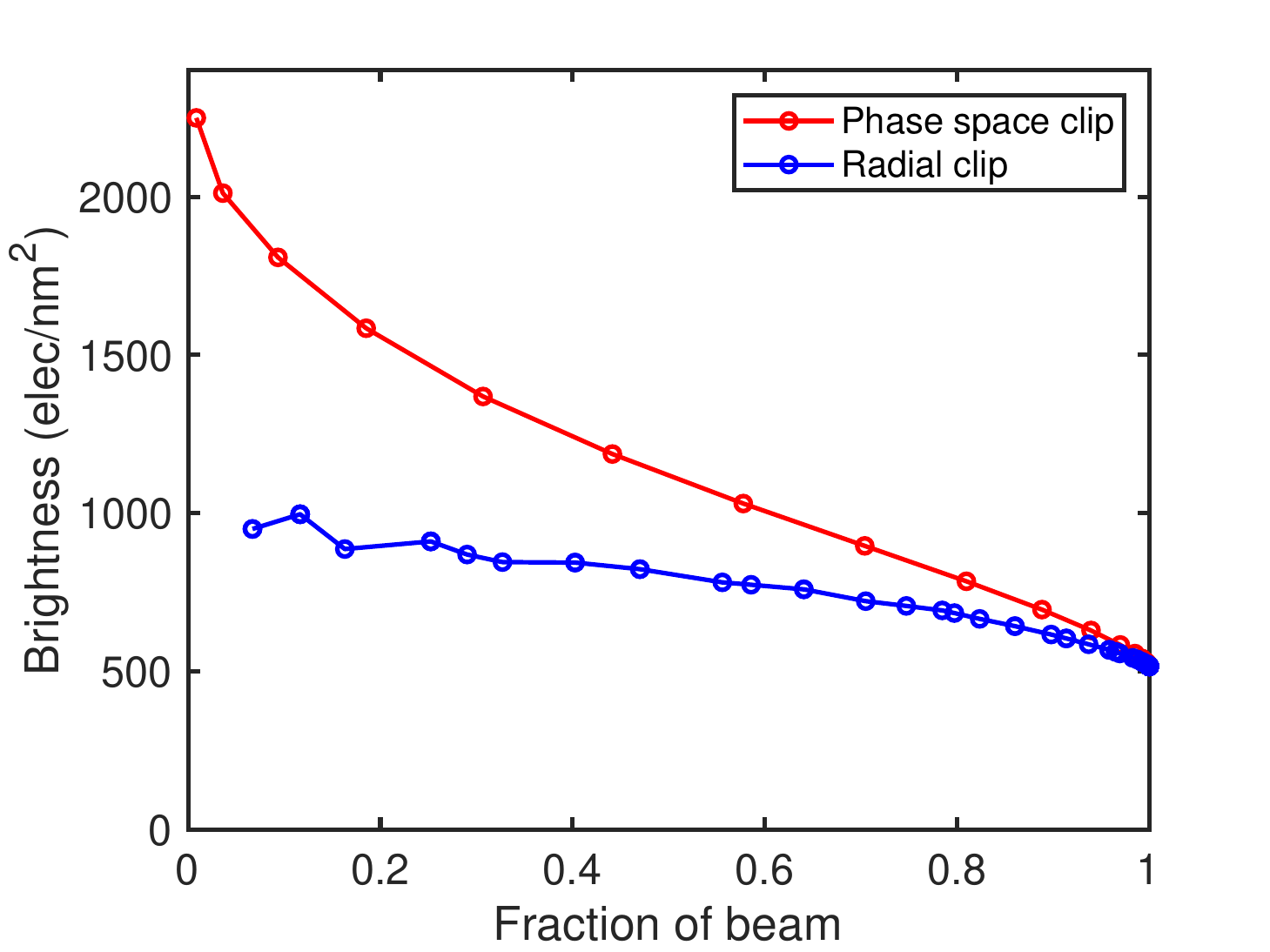}
\caption{Measured 4-d Brightness of a beam with a total charge 15 fC on the aperture as a function of the fraction of bunch charge included. In blue we restrict charge with a circular spatial aperture of varying size. In red we restrict charge with a shrinking phase space ellipse aligned to the beam's transverse phase space distribution, which would not be possible to achieve with a single physical aperture.  }  
\label{fig:bright}
\end{figure}

The impact of the the visible space-charge induced phase space distortions is partially mitigated by our use of a probe defining aperture, in that the outer spatial regions of the beam will be removed. Our 4-d phase space measurement procedure allows us to determine what the brightness benefit of such an aperture is. Fig. \ref{fig:bright} shows an analysis of a higher charge case with 15 fC incident on the aperture. The blue line shows the effects of a physical circular pinhole of variable radius. The four dimensional brightness increases from a full-beam value of $\sim 500$ electrons/nm$^2$ to a value around 1000 electrons/nm$^2$, in agreement with our previous measurements reported in Ref \cite{MEDUSA-Performance-Paper}. While a significant improvement, a pinhole merely clips in space, and the emittance growth from space charge does not occur all at one phase advance relative to the aperture. Naturally, we therefore do not transmit only the brightest portion of the phase space through a single aperture. For comparison, the red line of Fig. \ref{fig:bright} shows the brightness as we use a fictional elliptical phase space aperture in both the $x$ and $y$ phase spaces with ellipses aligned to the beam's phase space orientation. This procedure captures the brightest differential portion of phase space, with brightness here measured near 2200 electrons/nm$^2$. Of course, no single pinhole can select this region of phase space; in principle it might be possible with multiple pinholes with appropriate emittance preserving phase advance between them. 

\section{Summary and Outlook}
As photoinjectors for electron scattering and synchrotron radiation applications push for smaller emittance, aberrations which have previously been ignored will have increasing  impact on the downstream emittance of the electron beam. In our case, the search for emittance in the range of 10 nm for keV UED utilizing optics with high bunch charge and large convergence angles exposed effects from quadrupole and sextupole aberrations in our solenoids and rf cavity which would reduce brightness by orders of magnitude if left unaddressed. We utilized established correction techniques developed for erroneous quadrupole moments, and extended those to an experimental demonstration of simultaneous quadrupole and sextupole correction. 

Critical to this effort was the development of a precision 4-d phase-space mapping system which allowed us to measure not just emittance with sub-nm resolution, but to directly extract the correlations between the two transverse planes. Ultimately, after correction our measurements are consistent with simulations without aberrations, and we achieve very high brightness beams capable of new regimes in parameter space for ultrafast micro-diffraction \cite{MEDUSA-Performance-Paper}. Finally, we highlight the important role that spatial aperturing can play in mitigating emittance growth from space charge, which is a significant additional benefit beyond precisely defining the probe beam size in micro-diffraction experiments.   

\begin{acknowledgements}
 This work was supported by the U.S Department of Energy, grant DE-SC0020144 and U.S. National Science Foundation Grant PHY-1549132, the Center for Bright Beams.
 \end{acknowledgements}
 
M. Gordon and W. H. Li contributed equally to this work.

\end{document}